\documentstyle[12pt,twoside,fleqn,espcrc1]{article}

\newcommand{\AmS}{{\protect\the\textfont2
  A\kern-.1667em\lower.5ex\hbox{M}\kern-.125emS}}
\newcommand{\dst}{\displaystyle\strut}
\newcommand{\ov}{\over \displaystyle\strut}
\newcommand{\ben}{\begin{eqnarray}}
\newcommand{\enn}{\end{eqnarray}}
\def\l({\left(}
\def\r){\right)}
\bf

\title{Bose-Einstein Correlations for Expanding Finite Systems}

\author{T. Cs\"org\H o\address{ MTA KFKI RMKI,
	 H-1525 Budapest 114, P.O.B. 49, Hungary; \\
	Department of Physics,
        Columbia University, New York, USA }$^{\rm\scriptsize , b}$
        \thanks{Supported in part by grants OTKA-T2973, F4019 and
	W015107, MAKA 378/93 and Soros C5309/94}
        and
        B. L\"orstad\address{ Department of Physics, University of Lund, \\
        S\"olvegatan 14A, S-223 62 Lund, Sweden}}

\begin{document}
\rightline{CU-TP-684}
\rightline{LUNFD6/(NFFL-7106) 1995}
\rightline{hep-ph/9503494}
\vfill
\begin{center}
\Large\bf
Bose-Einstein Correlations \\
for Expanding Finite Systems
\end{center}
\medskip
\begin{center}
T. Cs\"org\H o$^{a,b}$ and B. L\"orstad$^b$
\end{center}

\begin{center}
\it
$^a$ MTA KFKI RMKI, \\
H-1525 Budapest 114, P.O.Box 49  \\
and \\
Department of Physics, \\
Columbia University, New York, NY 10027\\
\end{center}
\begin{center}
\it
$^b$ Department of Physics, University of Lund,\\
S\"olvegatan 14A, S-223 62 Lund, Sweden
\end{center}
\vfill
\begin{center}
{March 11, 1995}
\end{center}
\vfill
\begin{abstract}
{\small
{\bf Abstract:}
 There are {\it two length-scales}
present simultaneously in all the principal
directions for three-dimensionally expanding,
finite systems. These are discussed in detail
for the case of a longitudinally expanding system
with a transverse flow and a transverse temperature profile.
For systems with large geometrical sizes we find an
{\it $m_t$-scaling} for the parameters of the Bose-Einstein
correlation function, which may be valid
in the whole transverse mass region for certain model parameters.
In this limit, the Bose-Einstein correlations
view only  a small part of the source. The large geometrical
sizes can be inferred from a simultaneous analysis of the
invariant momentum distribution and the Bose-Einstein correlation
function. A preliminary analysis of the NA44 data indicates
that instead of a small fireball we are observing
a big and expanding {\it snowball} in $S + Pb $ reactions
at CERN SPS energies.
}
\end{abstract}
\vfill
\leftline{---------}
\leftline{\small
Contribution to the Quark Matter'95 conference,
Monterey, CA, USA, January 1995. \hfill}
\leftline{\small Nucl. Phys. A in press, ed. A. Poskanzer et al.\hfill}

\eject
\maketitle

\begin{abstract}
Bose-Einstein correlations and invariant momentum distributions
are presented for expanding finite systems
with applications to recent NA44 data.
\end{abstract}

\section{Introduction}

Bose-Einstein correlation functions  (BECF-s)
provide a unique tool for
the analysis of the freeze-out geometry on the fermi scale.
They are also subject to lots of non-ideal effects which make the
analysis and interpretation of these data rather difficult.

There are {\it two length-scales}
present  simultaneously
in all the principal directions of three dimensionally
expanding, finite systems~\cite{nr}. One of the length-scales is the
geometrical size $R_G$, the other is generated by the
freeze-out temperature distribution and the gradients of the flow.
This second kind of radius is referred to either  as
the 'thermal radius' $R_T$~\cite{nr,1d,3d} or the 'length of homogeneity'
{}~\cite{sinyukov,sinyukov2}. The thermal radius characterizes the size
of the region in the fluid from which particles with similar
momentum are emitted, while the geometrical radius is present due to
the finite size of the expanding
system.

\section{Model specification}
We model the emission function for high energy heavy ion reactions
as
\begin{eqnarray}
S(x,K) \, d^4 x & = & {\dst  g \ov (2 \pi)^3} \,  m_t \cosh(\eta - y) \,
\exp\l( - {\dst K \cdot u(x) \ov  T(x)} + {\dst \mu(x) \ov  T(x)}\r)
  \, H(\tau) d\tau  \, \tau_0 d\eta \, dr_x \, dr_y, \\
u(x) & \simeq & \l( \cosh(\eta)
	\l(1 + b^2 \, {\dst r_x^2 + r_y^2 \ov 2 \tau_0^2}\r) ,
	\,\, b \,{\dst r_x \ov \tau_0}, \,\, b \,{\dst r_y \ov  \tau_0}, \,\,
	\sinh(\eta)
	\l( 1 + b^2 \,{\dst r_x^2 + r_y^2 \ov 2 \tau_0^2} \r) \, \r),\\
T(x) & = & {\dst T \ov 1 + a^2 \, {\dst  r_x^2 + r_y^2 \ov 2 \tau_0^2} }
\qquad\mbox{\rm and} \qquad
{\dst \mu(x) \ov T(x) }  =  {\dst \mu_0 \ov T} -
	{ \dst r_x^2 + r_y^2 \ov 2 R_G^2}
	-{ \dst (\eta - y_0)^2 \ov 2 \Delta \eta^2 }.
\end{eqnarray}
Thus we include a finite duration,  $H(\tau) \propto \exp(-(\tau-\tau_0)^2
/(2 \Delta\tau^2))$.
The decrease of the temperature in the transverse direction is
controlled by the parameter $a$, while
the strength of the transverse flow is controlled by the parameter $b$.
A non-relativistic limit for $a = 0$ is described in ref.~\cite{nr}.
One dimensionally expaning finite systems correspond to the
$a = b = 0$  case~\cite{1d}, see this paper
for the notation too.
The first version of the three dimensionally expanding
cylindrically symmetric finite model
corresponds to the $a = 0$ and $b=1$ case~\cite{3d}.
 The integrals of the emission function  are evaluated
using the saddle-point method ~\cite{sinyukov,sinyukov2,uli}.
The saddle-point equations are solved in the LCMS~\cite{1d}, the longitudinally
comoving system,  for $\eta_s <<1 $ and $r_{x,s} << \tau_0 $.
These approximations are warranted if
$ \mid y - y_0 \mid << 1 + \Delta \eta^2 m_t /T$ and
 $\beta_t
 = p_t / m_t << (a^2 + b^2) / b$.
The flow is non-relativistic at the saddle-point if $\beta_t
  << (a^2 + b^2) / b^2$.
The radius parameters are evaluated here up to
${\cal O}(r_{x,s}/\tau_0) + {\cal O}(\eta_s)$, keeping only the leading order
terms in the LCMS. However, when evaluating the invariant momentum
distribution (IMD),
sub-leading terms coming from the $\cosh(\eta-y)$ pre-factor
are also summed up,
since this factor influences the IMD in the lower $m_t$ region where
the data are very accurate.

\subsection{Bose-Einstein correlations }
The BECF  is parameterized in the form of
$C(Q_L,Q_{side},Q_{out}) = 1 + \lambda \exp(-R_L^2 Q_L^2 - R_{side}^2
Q_{side}^2 - R_{out}^2 Q_{out}^2) $ where the intercept parameter
$\lambda$ and the radius parameters may depend on the rapidity
and the transverse mass of the pair.
We obtain~\cite{1d,3d}
\begin{eqnarray}
R_{side}^2 & = & R_*^2, \qquad
R_{out}^2   =  R_*^2 + \beta_t^2 \Delta \tau^2, \qquad
R_{L}^2  =  \tau_0^2 \Delta \eta_*^2,\\
{\dst 1 \ov R_*^2 } & = & {\dst 1 \ov R_T^2 } + {\dst 1 \ov R_G^2}
\qquad \mbox{\rm and} \qquad
{\dst 1 \ov \Delta \eta_*^2}  =  {\dst 1 \ov \Delta \eta^2 } +
		{\dst 1 \ov \Delta \eta_T^2} ,
\end{eqnarray}
i.e. the parameters of the BECF are dominated by the smaller of the
geometrical and the thermal length-scales. The
thermal length-scales $R_T$ and $\Delta\eta_T$
are found to be
\ben
 R_T^2 & = & {
\displaystyle\strut
\tau_0^2 \over
\displaystyle\strut
 a^2 + b^2 } {
\displaystyle\strut
T \over
\displaystyle\strut
 m_t}\qquad \mbox{\rm and} \qquad
\Delta\eta_T^2  =  {\dst T \ov m_t}.
\enn
These analytic expressions indicate that the BECF views only
part of an expanding source. Even a complete measurement of the
parameters of the BECF as a function of the mean momentum of the
particles may be insufficient to determine uniquely the underlying
space-time emission function due to this reason.
If the finite source sizes are large compared to the thermal
length-scales and if we also have $a^2 + b^2 \approx 1$,
one obtains an
$m_t$ -{\it scaling} for the parameters of the BECF,
\ben
R_{side}^2 &\simeq & R_{out}^2 \simeq R_L^2 \simeq \tau_0^2 {\dst T \ov m_t},
\quad \mbox{\rm valid for} \quad \beta_t << {\dst (a^2 + b^2) \ov b^2}
\simeq {1 \ov b^2}.
\enn
Note that this relation is independent of the particle type and has been
seen in NA44 data~\cite{na44}.
This $m_t$-scaling may be valid to arbitrarily large
transverse masses with $\beta_t \approx 1$ if $b^2 << 1$.
The finiteness of the expanding system reveals itself in an out-long
cross term too~\cite{uli} which is next to leading order in the LCMS,
being smaller than $R_{out}^2 - R_{side}^2$,
 which measured difference is very small ~\cite{na44},
alternatively explained by ref.~\cite{sudden}.

\subsection{Invariant momentum distributions}
The IMD plays a {\it complementary role}
to the measured Bose-Einstein correlation function ~\cite{nr,1d,3d}.
Namely, the width of the rapidity distribution at a given $m_t$
as well as
$ T_*$ the effective temperature at a mid-rapidity $y_0$ shall be
dominated by the {\it longer} of the thermal and geometrical
length-scales.
Thus a {\it simultaneous analysis} of the Bose-Einstein correlation function
and the IMD may reveal information
both on  the temperature and flow profiles  and on the geometrical sizes.
E.g. the following relations hold:
\begin{eqnarray}
\Delta y(m_t)^2 & = & \Delta \eta^2 + \Delta \eta_T^2(m_t), \qquad
\mbox{\rm and} \qquad
{\dst 1 \ov T_*}  =  {\dst f \ov T + T_G(m_t=m) } + {\dst 1 -f \ov T}.
\end{eqnarray}
The geometrical contribution to the effective temperature is given by
$T_G  = T \,  R_G^2 / R_T^2$ and the fraction $f$ is defined as
$f = b^2 / (a^2 + b^2)$, satisfying $0 \le f \le 1$.
The saddle-point sits at $\eta_s =
(y_0 - y) / (1 + \Delta\eta^2 / \Delta \eta_T^2 )$, $r_{x,s} =
\beta_t b R_*^2 / (\tau_0 \Delta\eta_T^2)$ and  $r_{y,s} = 0$.

For the considered model, the invariant momentum distribution
can be calculated as
\ben
{\dst d^2 n \ov dy \, dm_t^2 } & = &
	{\dst g \ov (2 \pi)^3 } \,\, \exp\l( {\dst \mu_0 \ov T} \r) \, \,
	m_t \,\, (2 \pi \Delta\eta_*^2 \tau_0^2)^{1/2} \, \,
	(2 \pi R_*^2) \,\, \cosh(\eta_s) \, \,
	\exp(+ \Delta \eta_*^2 / 2) \times  \nonumber \\
\null & \null & \times \exp\l( - {\dst (y - y_0)^2 \ov 2 (\Delta\eta^2 +
	\Delta \eta_T^2) } \r) \,
\exp\l( - {\dst m_t \ov T} \l(  1 - f \, {\dst \beta_t^2 \ov 2} \r) \r) \,
\exp\l( - f \, {\dst m_t \beta_t^2 \ov 2 (T+ T_G)} \r).
\label{e:imd}
\enn
This  IMD has a rich structure:
it features { both a rapidity-independent
and  a rapidity-dependent low-pt enhancement}
as well as a
{ high-pt enhancement or decrease}.

Within this model,
the {\it rapidity-independent low-pt enhancement} is a consequence
of the transverse mass dependence of the effective volume,
from wich particles with a given momentum are emitted.
This is described by the $ (2 \pi \Delta\eta_*^2 \tau_0^2)^{1/2}
(2 \pi R_*^2) \cosh(\eta_s) \exp(+ \Delta\eta_*^2 / 2) $ factor,
being proportional to $ (T/m_t)^{(3/2)} \exp(+ T /(2 m_t) )$.

The {\it rapidity-dependent low-pt enhancement}, which is a generic
property of the longitudinally expanding finite systems~\cite{lpte},
reveals itself in the rapidity-dependence of the effective temperature,
defined as the slope of the exponential factors in the IMD
 in the low-pt limit at a given value of the rapidity.
The leading order ~\cite{lpte} result is
\ben
T_{eff}(y) & = & {\dst T_* \ov 1 + a (y - y_0)^2 }
\qquad \mbox{\rm with} \qquad a = {\dst T T_* \ov 2 m^2}
	\l(\Delta\eta^2 + {\dst T \ov m}\r)^{-2}.
\enn

The {\it high-pt enhancement or decrease}
refers to the change
of the
effective temperature at mid-rapidity with increasing $m_t$ .
The large transverse mass limit
$T_{\infty}$  shall be in general
different from the effective temperature at low pt given by $T_*$
since
\ben
T_{\infty} & = & {\dst 2 T \ov 2 - f} \qquad \mbox{\rm and}\qquad
{\dst T_{\infty} \ov T_* }  =  {\dst 2 \ov 2 - f} \l( 1 - f {\dst T_G(m)
	\ov T + T_G(m) }\r).
\enn
Utilizing $T_G / T = R_G^2 / R_T^2$, the
high-pt enhancement or decrease turns out to be controlled by the
ratio of the thermal radius $R_T(m_t = m)$ to the geometrical radius $R_G$.
One obtains $T_{\infty} > T_* $ if $ R_T^2(m) > R_G^2$
and similarly $T_{\infty} < T_* $ if $ R_T^2(m) < R_G^2$.
Since for large colliding nuclei $R_G$ is expected to increase,
the high-pt decrease in these reactions becomes a geometrical
effect, a consequence of the large size.

\section{Large Halo: Applications to NA44 data}
The effects of long-lived resonances on both the IMD and the BECF
can be taken into account analytically along the lines of ~\cite{halo},
supposing that the halo is sufficiently large.
NA44 data~\cite{na44} for central $S + Pb$ reactions at CERN SPS
with
200 AGeV show an approximately $m_t$ independent  intercept
parameter:
$\lambda_{\pi^+} = 0.56 \pm 0.02$ and $0.55 \pm 0.02$ at
$ m_t = 150 $ MeV and $ 450 $ MeV, respectively. This suggests that
the halo contains $ 1 - \sqrt{\lambda}
=
25 \pm 2 $ \% of all the pions.
If $\lambda = const$
the equations simplify~\cite{halo}  as
\ben
{\dst d^2 n \ov dy \, dm_t^2 } \, & = & \, {\dst d^2 n_c \ov dy \, dm_t^2 }
                                \, = \, {\dst d^2 n_h \ov dy \, dm_t^2 },
\qquad \mbox{\rm and} \qquad
{ C(\Delta k, K) } \,   =  \, 1 + \lambda R_c(\Delta k, K).
\enn
In this case
 the only apparent effect of the halo is to reduce the intercept
parameter of the measured BECF to $\lambda  = const $ while the IMD and
the $(\Delta k, K)$ dependence of the BECF is determined by the central
part apparently {\it exclusively}.
This central part is well accessible to the analytical models,
presented e.g. in the previous sections.
In general the halo may have more influence both on the IMD and the BECF.

In a preliminary analysis, we have fitted the NA44 preliminary
IMD for pions and kaons together
with the final NA44 data~\cite{na44} for the $m_t$ dependence of the
BECF parameters for both pions and kaons.
Fixing the parameters $y_0 = 3$, $a = 0$ and $b = 1$
and extrapolating the equations to the relativistic $m_t$ domain
we get a description of the IMD and the BECF for both pions and kaons
at a $\chi^2 / NDF = 2.2$. The preliminary analysis indicates
large geometrical source sizes for the pions, $R_G(\pi) \approx 8$ fm,
late freeze-out times of $\tau_0 \approx 8 $ fm/c, a vanishing duration
$\Delta \tau$ for both pions and kaons and finally surprisingly
low freeze-out temperatures, $T \approx 80 $ MeV.
The kaons appear from a smaller region of $R_G(K) \approx 3.5$ fm.

\section{Conclusions:} Instead of observing a small fireball
we are observing a big and expanding {\it snowball}
at CERN SPS $S + Pb$ reactions accoringly to the above
preliminary analysis of
the partly preliminary NA44 data.
The snowball may contain a hot core,
to be investigated in a future data analysis.
The preliminary results are very sensitive to the
detailed structure of the $m_t$-dependence of the
parameters of the Bose-Einstein correlation functions as well
as to the details of the invariant momentum distribution
of both pions and kaons.


\end{document}